\documentclass[prl,twocolumn,showpacs,superscriptaddress]{revtex4-1}
\usepackage{graphicx}
\usepackage{dcolumn}

\usepackage{lineno}
\usepackage{eucal}
\usepackage[dvips]{epsfig}
\usepackage{amssymb}

\usepackage{amsmath,amsthm}
\usepackage{color}

\usepackage{mathtools}

\newcommand{\be}{\begin{eqnarray}}
\newcommand{\ee}{\end{eqnarray}}
\newcommand{\bse}{\begin{subequations}}
\newcommand{\ese}{\end{subequations}}

\newcommand{\bnum}{\begin{enumerate}}
\newcommand{\enum}{\end{enumerate}}

\newcommand{\bit}{\begin{itemize}}
\newcommand{\eit}{\end{itemize}}

\newcommand{\bc}{\begin{cases}}
\newcommand{\ec}{\end{cases}}

\newcommand{\bpm}{\begin{pmatrix}}
\newcommand{\epm}{\end{pmatrix}}

\newcommand{\bvm}{\begin{vmatrix}}
\newcommand{\evm}{\end{vmatrix}}

\newcommand{\bs}{\boldsymbol}

\newcommand{\mcal}{\mathcal}

\newcommand{\ga}{\alpha}
\newcommand{\gb}{\beta}

\newcommand{\eps}{\epsilon}

\newcommand{\gl}{\lambda}

\newcommand{\gr}{\rho}

\newcommand{\f}{\frac}

\newcommand{\csp}{\;,\qquad}

\begin{document}
\title{Hydrodynamic length-scale selection and effective viscosity in 
microswimmer suspensions}

\author{Sebastian Heidenreich}
\email{sebastian.heidenreich@ptb.de}
\affiliation{Department of Mathematical Modelling and Data Analysis,
Physikalisch-Technische Bundesanstalt Braunschweig und Berlin, Abbestr. 2-12,
10587 Berlin, Germany}

\author{J\"orn Dunkel}
\affiliation{Department of Mathematics, Massachusetts Institute of Technology,
77 Massachusetts Avenue E17-412, Cambridge, MA 02139-4307}

\author{Sabine H. L. Klapp}
\affiliation{Institute for Theoretical Physics, Technical University Berlin,
Hardenbergstr. 36, D-10623, Berlin, Germany} 

\author{Markus B\"ar}
\affiliation{Department of Mathematical Modelling and Data Analysis,
Physikalisch-Technische Bundesanstalt Braunschweig und Berlin, Abbestr. 2-12,
10587 Berlin, Germany}

\date{\today}

\begin{abstract} 
A universal characteristic of mesoscale turbulence in active suspensions is
the emergence of a typical vortex length scale, distinctly different from the
scale-invariance of turbulent high-Reynolds number flows. Collective
length-scale selection has been observed in bacterial fluids, endothelial tissue and active colloides, yet the physical origins of this phenomenon remain elusive. Here, we
systematically derive an effective fourth-order field theory from a generic
microscopic model that allows us to predict the typical vortex size in
microswimmer suspensions. Building on a self-consistent closure condition, the  
derivation shows that the vortex length scale is determined by the competition
between local alignment forces and intermediate-range hydrodynamic interactions.
Vortex structures found in simulations  of the theory agree with recent
measurements in {\em Bacillus
subtilis} suspensions. Moreover, our approach correctly predicts an effective
viscosity enhancement (reduction), as reported experimentally for puller (pusher)
microorganisms.  
\end{abstract}


\maketitle
A universal feature shared by many living systems is the emergence of
characteristic length and time scales that arise from the non-equilibrium
dynamics of their microscopic constituents.  Examples range from circadian
oscillations in individual cells~\cite{1999Dunlap} to multicellular
gene-expression patterns in embryos~\cite{2006DevelopmentalBiology_Book} and
vortex structures in microbial
suspensions, endothelial tissue and active colloides~\cite{Sano2015, Rossen2014,
PNAS,zhang2009swarming}. Yet, despite their
broad biological relevance, it has proved difficult to predict quantitatively
how such emergent scales arise from the underlying chemical or physical
parameters.  In the past decade, bacterial and other active
suspensions~\cite{PNAS, Aranson_collective_motion, Rossen2014}
have emerged as important biophysical model systems that can help bridge the gap
between large-scale spatio-temporal pattern formation and microscopic
non-equilibrium dynamics~\cite{SoftMatter_Review}. At high densities, bacterial
fluids 
form coherent vortex structures, spanning several cell lengths in
diameter~\cite{Aranson_collective_motion,PNAS, PRL}  and persisting for several
seconds~\cite{PRL} or even minutes~\cite{2013Wioland_PRL,Lushi14,
Grossmann1}.  Although a number of insightful theoretical models have been
proposed~\cite{MarchettiP, Peshkov, MarchettiQ,Ramaswamy, PNAS_Baska}, a
quantitative theory connecting microswimmer properties to the experimentally
observed vortex patterns has been lacking. 
\par
Here, we present such a theory by drawing guidance from the recent
observation~\cite{PNAS, PRL} that an effective fourth-order continuum model can
provide a quantitative phenomenological description of dense bacterial
suspensions~\cite{NJP}. This model, which combines the seminal Toner-Tu
description
of flocking~\cite{Toner_Tu}  with the Swift-Hohenberg equation from pattern
formation \cite{SwiftHohenberg}, describes the effective bacterial velocity
field ${\bf w}(t,\bf x)$ by 
\begin{eqnarray}
\notag (\partial_t  - \lambda_0 {\bf w}
\cdot \nabla  ){\bf w} &=&
 -\nabla q + \lambda_1 \nabla
|{\bf w}|^2 + \alpha {\bf w} - \beta |{\bf w}|^2 {\bf w}
\\
&&
+ \Gamma_0  \nabla^2 {\bf w}    - \Gamma_2  (\nabla^2)^2 {\bf w},
\label{DH}
\end{eqnarray}
where the bacterial pressure field $q(t,\bf x)$ accounts for incompressibility,
$\nabla\cdot{\bf w} =0$. Although a direct fit of Eq.~(\ref{DH}) can reproduce
the key statistical features of experimental data for dense quasi-2D~\cite{PNAS}
and 3D~\cite{PRL} {\em B. subtilis} suspensions, the connection between the
phenomenological parameters $(\gl_0,\gl_1,\ga,\gb,\Gamma_0,\Gamma_2)$ and
individual bacterial properties has remained unknown. Below, we systematically
derive a generalized variant of Eq.~\eqref{DH} directly from a generic
microswimmer model. The derivation specifies each parameter in the continuum
theory in terms of microscopic swimmer parameters and yields direct 
theoretical predictions for the typical vortex size and effective viscosity in dense microswimmer
suspensions.  Compared with previous studies, our approach differs in that we
deduce a self-consistent closure condition that accounts for shear-induced tumbling, a
physically important effect that has previously been neglected. We present a
bifurcation analysis of the resulting 
fourth-order model and discuss numerical results demonstrating satisfactory
agreement with available experimental data for quasi-2D
suspensions~\cite{Aranson_collective_motion}.

\par
\paragraph{Microscopic model.--}
We consider microswimmers moving in an incompressible Newtonian fluid at low
Reynolds number, described by the Stokes equations
\be
\label{SE} 
-{\nabla}  p + \mu \nabla^2{\bf u} + 
{\nabla} \cdot {\boldsymbol \sigma}=0,
\qquad \nabla\cdot {\bf u}=0.
\ee 
Here, ${\bf u}(t,\bf x)$ is the fluid velocity, $p(t,\bf x)$ the hydrodynamic
pressure and  $\mu$ the effective  dynamic viscosity.  The active stress tensor
$ {\boldsymbol \sigma}$ represents the forcing of the fluid by the
swimmers~\cite{PNAS_Baska, Ramaswamy,SoftMatter_Review}.
For dense suspensions,  the bulk viscosity $\mu$ contains contributions from the
solvent as well as passive and active contributions from the
microswimmers~\cite{Heines08,Sokolov09,PhysRevLett.110.268103,PhysRevLett.104.098102,2011Ryan_PRE,
Gluzman:2013aa}.  For simplicity, we assume that the passive contribution is
approximately given by the Batchelor-Einstein relation for spherical colloids,
$\mu =  \mu_0 \left(1 + k_1 \phi  + k_3 \phi^2\right)$, where $\mu_0$ is the \lq
bare\rq{} solvent viscosity,  $\phi$ is the volume fraction and $k_i$ are
positive constants~\cite{Hinch10, Heines08, Batchelor, Einstein}.   The active
contribution will be derived below. In quasi-2D Hele-Shaw flow, an effective
boundary-friction term $-\nu_0\bf u$ is added~\cite{2013Woodhouse_PNAS} on the
lhs. of the Stokes equation~\eqref{SE}. 
\begin{figure}
\begin{center} 
\includegraphics[width=\columnwidth]{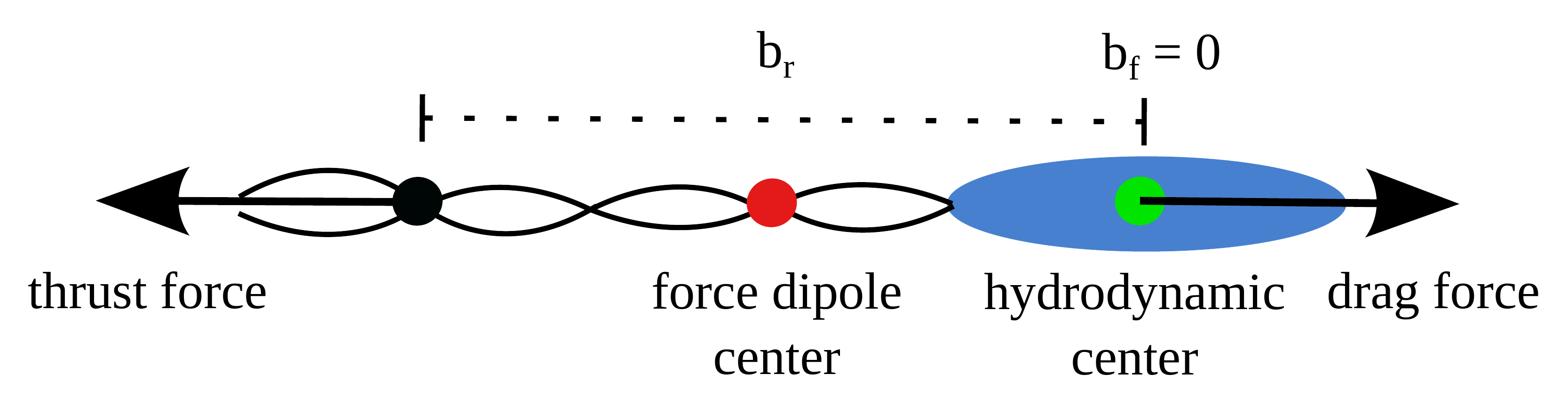} 
\caption{\label{sketch} 
(color online) Schematic of a bacterial microswimmer such as {\em
B. subtilis}. The center of the hydrodynamic stress $\bs X$ (green) is located in front of the force dipole  center (red) ~\cite{Knut}. $b_r$ and
$b_f$ are the distances between the center $\bs X$ and acting forces. The drag force defines the swimmer orientation $\bs N$.}
\end{center}
\end{figure}
\par
Focussing on time scales larger than a typical stroke period, we
describe microswimmers as force dipoles of strength $f_0$~\cite{Knut}
(Fig.~\ref{sketch}).  Assuming $\sigma=1,\ldots, M$ identical swimmers, their
time-dependent positions ${\bf X}^\sigma(t)$ and orientation unit vectors ${\bf
N}^\sigma(t)$ are determined by the overdamped Langevin
equations~\cite{Shelley1, Shelley2}
\begin{eqnarray}
\label{LE}
\dot{\bf X}^\sigma  &=& v_0  {\bf N}^\sigma +  {\bf u}(t,{\bf X^\sigma}) +
\sqrt{2 D}
\boldsymbol{\xi}^\sigma,  \\
\label{LE1}
\dot{\bf N}^\sigma &=& {\boldsymbol \Pi}^\sigma \cdot \left[
\nabla {\bf u}\cdot {\bf N}^\sigma   - 
{\boldsymbol
\nabla}_{{\bf N}^\sigma} \Phi  +
\frac{1}{\sqrt{\tau}}{\boldsymbol \eta}^\sigma  \right], 
\end{eqnarray}
where overdots indicate time-derivatives. The translational dynamics~\eqref{LE}
is caused by self-swimming at speed~$v_0$, hydrodynamic advection~${\bf
u}(t,{\bf X^\sigma})$ and translational Brownian motion of strength $D$. The
random functions  $\boldsymbol{\xi}^\sigma(t)$ and $\boldsymbol{\eta}^\sigma(t)$
 denote independent $\delta$-correlated Gaussian white noise. The orientational
dynamics~\eqref{LE1}, interpreted as a Stratonovich stochastic differential
equation~\cite{Jacobs} with rotational relaxation time $\tau$, conserves the
length of the orientation vector $\bf N^\sigma$ by virtue of the projector
${\boldsymbol \Pi}^\sigma ={\bf{I}} -{\bf N}^\sigma {\bf  N}^\sigma$, 
where~${\bf I}$ is the unit matrix. The $\nabla {\bf u}$-term accounts for
reorientation of elongated particles by flow
gradients~\cite{Jeffery,HoheneggerShelley}. In dense suspensions of
fast-swimming bacteria, steric collisions are negligible in the translation
dynamics but may contribute significantly to reorientation. We 
therefore include a polar reorientation interaction potential $\Phi( {\bf
N}^\sigma) =-g \sum_{|{\bf X}^\sigma - {\bf X}^\nu |\le \epsilon} {\bf N}^\sigma
\cdot {\bf N}^\nu$ with cut-off length $\epsilon$ and alignment strength $g =
g_0 v_0/2$ in Eq.~\eqref{LE1}, reflecting the experimental  observation of
locally aligned bacterial jets~\cite{Swinney,zhang2009swarming,PNAS}.  In
particular, we assume
that kinematically induced polar interactions dominates over nematic
ordering~\cite{2006Peruani}, the latter representing the dominant alignment
force in the passive limit~$v_0\to 0$.  
\par
Although Eqs.~\eqref{LE} and~\eqref{LE1} as well as the main steps of the
subsequent derivation remain valid for 3D bulk suspensions, we  focus, for
clarity, on free-standing quasi-2D
films~\cite{Aranson_collective_motion,2010Gollub_PRL,2011Gollub_PNAS} from now
on.

\par
\paragraph{Fokker-Planck dynamics.--}
To derive a continuum model from Eqs.~\eqref{SE}-\eqref{LE1}, we consider the
one-particle distribution  \mbox{$\mathcal{P}(t,{\bf x },{\bf n}) =
M^{-1}\sum_{\sigma = 1}^M \langle \delta({\bf x} - {\bf X}^\sigma) \delta({\bf
n} - {\bf N}^\sigma) \rangle$}, where $\langle \,\cdot\, \rangle$ denotes an
average over the Gaussian white
noise~$\{\boldsymbol{\xi}^\sigma(t),\boldsymbol{\eta}^\sigma(t)\}$. The
evolution of $\mathcal{P}(t,{\bf x },{\bf n})$ is governed by the Fokker-Planck
equation~\cite{Risken, Jacobs}
\begin{eqnarray}
\label{FP}
&&\partial_t \mathcal{P} =  - {\nabla}\cdot (v_0 {\bf n} + {\bf u})  \mathcal{P}
+ D {\nabla}^2\mathcal{P}
  -  {\nabla}_{\bf n} \cdot {\boldsymbol \Pi} \cdot
({\nabla} {\bf u}) \cdot {\bf n} \; \mathcal{P} 
 \notag \\
&& +     \frac{1}{\tau} {\nabla_{\bf n}} \cdot {\bf  n} \;
\mathcal{P}    + \frac{1}{2 \tau} {\nabla}_{\bf n} {\nabla}_{\bf n}
:({\boldsymbol \Pi}\cdot {\boldsymbol \Pi}^T )\mathcal{P} 
  + \mathcal{C}^{(2)}\left[\Phi \right].
\end{eqnarray}
Alignment interactions enter via the collision integral
\begin{equation}
\label{II}
\mathcal{C}^{(2)}[\Phi] =  
{\nabla}_{\bf n} \cdot 
\int d{\bf n}'\int_\epsilon d{\bf
x}'\; {\boldsymbol \Pi} \cdot  [{\nabla}_{{\bf n}} \Phi ({\bf
n},{\bf n}')]\;  \mathcal{P}^{(2)} ,
\end{equation}
which involves the two-particle distribution function $
\mathcal{P}^{(2)}(t,{\bf x},{\bf n};{\bf x}',{\bf n}')$. 

\par
\paragraph{Moment equations.--}
To derive hydrodynamic field equations from Eqs.~\eqref{FP}, and \eqref{II} we define the 2D
swimmer number density $\rho(t,{\bf x})=M\int d\bf n\, \mcal{P}$, the polar
order-parameter  field 
${\bf P}(t,{\bf x}) = \rho \overline {{\bf n}}$ and the nematic order-parameter
field ${\bf Q}(t,{\bf x}) =\rho (\overline{{\bf  nn} } - {\bf I}/2)$, where the
bar denotes the marginal average over all orientations~$\overline{\bf g }(t,{\bf
x}) =\int d{\bf n}\;  {\bf g}({\bf n})\mathcal{P}$.  Integrating  Eq.~\eqref{FP}
over $\bf n$ and considering the limit of constant density $\gr$ implies an 
incompressibility condition for the orientation field,~$\nabla\cdot{\bf P}=0$.
\par
To obtain the dynamic equation for ${\bf P}$, we multiply  Eq.~\eqref{FP} by $
{\bf n}$ and integrate over all orientations.  Adopting a standard mean-field
approximation for Eq.~\eqref{II}, assuming again constant density $\gr$ and
neglecting terms of tensorial rank higher than two, we find 
\begin{eqnarray}
\label{FE0}
&&\partial_t  {\bf  P} +  {\bf u} \cdot \nabla {\bf P} = 
-\nabla \hat{p}-v_0 \nabla \cdot {\bf Q} 
- (1/2) \, {\bf P} \cdot {\bf \Sigma} \notag  \\
&& \qquad
+ (D + D_\eps) \nabla^2 {\bf P} +  (\eps^2/24)D_\eps\nabla^4 {\bf P } \\
&& \qquad
+
\left[ \nabla {\bf u} + (4/\eps^2)D_\eps({\bf I}-2{\bf Q}) -(1/\tau){\bf I} 
\right] \cdot {\bf P} , \notag
\end{eqnarray}
where $\nabla^4= (\nabla^2)^2$. The local Lagrange multiplier $\hat{p}(t,{\bf
x})$ ensures a divergence-free orientation field, ${\boldsymbol \Sigma}
=\frac{1}{2}[{\nabla}{\bf u} +({\nabla}{\bf u})^\top ]$ is the hydrodynamic
rate-of-strain tensor,  and \mbox{$D_\eps=\rho g_0 v_0 \pi \epsilon^4/8$}
encodes translational diffusion caused by mean-field polar interactions. The
destabilizing $(\eps^2/24) D_\eps\nabla^4 {\bf P}$-term, arising from polar
interactions, is counteracted by the $ \nabla \cdot {\bf Q} $-term.
\par
To connect with Eq.~\eqref{DH},  we introduce an non-conserved self-swimming
velocity field $ {\bf v}(t,{\bf x}) = v_0 {\bf P}$.   The total velocity field
of the microswimmers, appearing in Eq.~\eqref{DH} and measured in
experiments~\cite{Aranson_collective_motion,PNAS, PRL}, is given by ${\bf w}=
{\bf u} + {\bf v}$, corresponding to the hydrodynamic average of Eq.~\eqref{LE}.

 \paragraph{Self-consistent closure.--}
To close Eq.~\eqref{FE0}, we have to approximate the nematic order-tensor ${\bf
Q}$ in terms of ${\bf P}$ and ${\bf u}$. A simple closure condition for passive
hard rods~\cite{Doi} is ${\bf Q} \sim ({\bf PP})^+$, where  ${\bf A}^+$ with
components 
$A_{ij}^+ =  (A_{ij} + A_{ji} - \delta_{ij} A_{k k})/2 $ denotes the
symmetric traceless part of tensor ${\bf A}$ in 2D.  However, this commonly used
closure conditions 
does not  account for the fact that active microswimmers permanently impose stress
on the ambient fluid which feeds back into the orientational order, analogous to the shear-induced isotropic-nematic transition in liquid
crystals~\cite{Hess75,Noirez,Callaghan,Kim}.
To derive a self-consistent closure that accounts for this effect, we multiply
Eq.~\eqref{FP} by $ {\bf n\bf n}$ and then integrate over all orientations. 
Taking the stationary limit of the resulting 
equation, assuming small spatial variations of {\bf Q} and neglecting \lq
flexoelectric\rq{}  contributions \cite{PRL_Nano}, one finds
\begin{equation}
\label{EQ}
{\bf Q} = \gl_\mathrm{P} ({\bf PP})^+ + \lambda_\Sigma\boldsymbol \Sigma,
\end{equation}
with \mbox{$\gl_\mathrm{P}=(2/\eps^2)D_\eps\tau$} and tumbling parameter~
\mbox{$\lambda_\Sigma= \tau/16$}.

\paragraph{Hydrodynamic stress.--}
As last step, we have to relate the active stress tensor $\boldsymbol \sigma$ in
Eq.~\eqref{SE} to ${\bf P}$. To this end, note that the swimmer position ${\bf
X}^\sigma$ coincides with the center of hydrodynamic stress (Fig.~\ref{sketch}), which is the point
where the hydrodynamic net torque on a rigid body vanishes~\cite{HappelBrenner}.
For an asymmetric dipole swimmer of effective length $\ell = b_f +b_r$ with
$b_f,b_r>0$, the propulsive rear force ${\bf F}^\sigma_r=-f_0 {\bf N}^\sigma$
acts at ${\bf X}_r^\sigma = {\bf X}^\sigma - b_r {\bf N}^\sigma$, and the
resistive front force  ${\bf F}^\sigma_f= f_0 {\bf N}^\sigma$ at ${\bf
X}_f^\sigma = {\bf X}^\sigma + b_f {\bf N}^\sigma$ (Fig. \ref{sketch}). The
swimmer is force-free,  ${\bf F}^\sigma_r + {\bf F}^\sigma_f=0$,  and for
bacterial pusher dipoles  we have $f_0>0$~\cite{Knut}. The total force density
is then given by 
\begin{equation}
 {\bf f} = \nabla \cdot \boldsymbol \sigma = \sum_\sigma {\bf F}^\sigma_r
\delta({\bf x} - {\bf X}_r^\sigma) + {\bf F}_f^\sigma \delta({\bf x} - {\bf
X}_f^\sigma).
\end{equation}
Taylor-expanding for small $b_r$ and $b_f$, assuming constant density across a
thin film of thickness $h\sim \ell$, neglecting terms of order
$(\nabla{\bf P})^2$, and averaging over noise and orientations~\cite{Ramaswamy,
PNAS_Baska}, we find for the divergence of the averaged total stress tensor~\footnote{To simplify subsequent equations, we drop $\mcal{O}(b_{r/f}^3)$-terms but kept the $\gamma_4$-term to ensure  stability.}
\begin{equation}
\label{actstress}
\nabla\cdot \overline{\boldsymbol \sigma} 
= 
-\f{f_0\rho}{h}
\nabla\cdot \left[
\ell {\bf Q} + 2\left(\gamma_2 +\gamma_4 \nabla^2\right)\left(\nabla
{\bf P}\right)^+
\right]
\end{equation}
with $\gamma_2 = (b_r^2 -b_f^2)/8>0$ and $\gamma_4 =(b_r^4 - b_f^4)/192>0$ for
pushers (in 3D, $\rho/h$ is replaced by the concentration~$c$).  Combining Eqs.~\eqref{EQ} and~\eqref{actstress} with~\eqref{SE} and \eqref{FE0},
we obtain a closed set of equations for the two incompressible
vector fields $({\bf u},{\bf P})$ and their associated pressure fields.

\begin{figure}
\begin{center}
\includegraphics[width=\columnwidth]{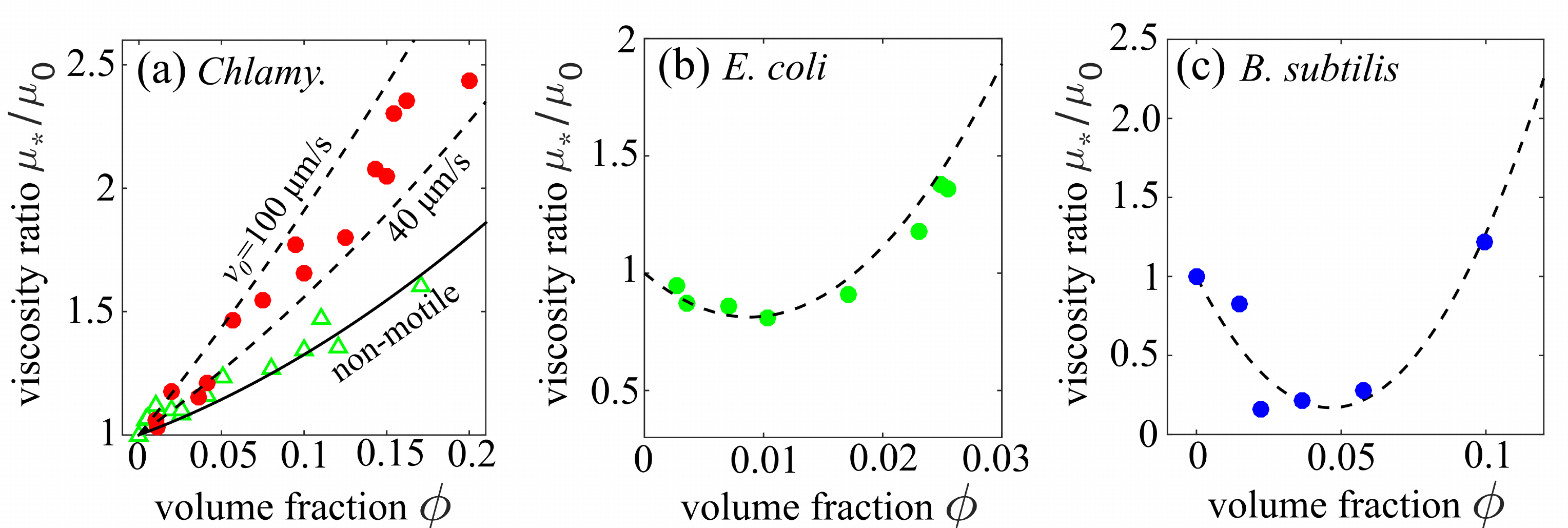} 
\caption{(color online) Comparison of predictions (dashed lines) based on
Eq.~\eqref{e:viscosity_result} with recent experiments (symbols) for force
density values $f_0$ given in the text. (a)~For
\textit{Chlamydomonas} algae (puller), the theory predicts an increase in the
viscosity for motile cells (dashed lines, using $\ell = 5 \,\mu$m, $\tau =
3.5$\,s, $k_2=7.6$ \cite{Hinch10, Heines08}) compared with non-motile cells
(solid line), in agreement with recent measurements for motile (red circles) and dead (green triangles) cells~\cite{PhysRevLett.104.098102}.  (b,c)  For
bacteria (pusher), the theory correctly predicts a  viscosity decrease  at
intermediate filling fractions. Dashed curves are based on the following fit parameters: (b) For    
\textit{E.~coli} data (green circles \cite{PhysRevLett.110.268103}): $v_0 = 20\,\mu$m/
s, $V_b = 1.57\, \mu$m$^3$, $k_3 = 2400$, $\tau = 4.8$s. (c) For 
\textit{B.~subtilis}  data (blue circles \cite{Sokolov09}): $v_0 = 30\, \mu$m/s, $V_b=1.92\, \mu$m$^3$, $k_3 =  385,\,\tau = 0.5$\,s.
\label{viscosity}
}
\end{center}
\end{figure}

\paragraph{Stokes equations \& viscosity.--}
Inserting Eq.~\eqref{actstress} into Eq.~\eqref{SE}, the Stokes equation can be
written as
\be
\label{e:stokes_new}
\mu_* 
 \nabla^2 {\bf u}
 -{\nabla}  p_*
=
f_0 \gr\left(\ell \gl_\mathrm{P} {\bf P}\cdot\nabla+
{\gamma_2} \nabla^2+
{\gamma_4}\nabla^4\right)
{\bf P},
\qquad
\ee
with effective viscosity and pressure given by
\be
{\mu}_* = \mu - \f{\ell \lambda_\Sigma}{2} \f{f_0\gr}{h},
\qquad
{p}_* =p -\f{\ell \gl_\mathrm{P}}{2} |{\bf P}|^2 \f{f_0\gr}{h}. 
\quad
\ee
Rewriting ${\mu}_*$ in terms of the volume fraction $\phi=\rho A,$ where $A$ is
the projected 2D area of a swimmer and choosing $h = \ell$, we find
\be
\label{e:viscosity_result}
\mu_* = \mu_0[1+(k_1-k_2)\phi + k_3\phi^2], 
\ee
where $k_1= 5/2 $ for passive spherical objects~\cite{Hinch10,
Heines08, Batchelor, Einstein} and $ k_2 =  f_0  \lambda_\Sigma (2\mu_0 A )^{-1}$. Thus, our theory implies
that, to linear order in $\phi$, pushers with $f_0>0$ can reduce the viscosity
whereas pullers with $f_0<0$ generally enhance the viscosity. In 3D, an
analogous derivation yields $k_2 =  f_0 \ell \lambda_\Sigma (2\mu_0 V_b )^{-1}$
where $V_b$ is the effective volume of the individual swimmer. These predictions
agree with recent measurements in 3D bacterial~\cite{Sokolov09,
PhysRevLett.110.268103} and algal~\cite{PhysRevLett.104.098102} suspensions
(Fig.~\ref{viscosity}).

\paragraph{Orientation field dynamics.--}
Taking the divergence of Eq.~\eqref{EQ} and exploiting the Stokes
equation~\eqref{e:stokes_new} yields~$\nabla\cdot{\bf Q}$ in terms of the
orientation field~${\bf P}$. Substituting the resulting expression into
Eq.~\eqref{FE0}, we find the fourth-order PDE
\be
\label{FEQ}
&&\partial_t  {\bf  P} + ( {\bf u}+v_0\gl_\mathrm{P*} {\bf P}) \cdot \nabla {\bf
P} 
= \notag
-\nabla \hat{p}_* - (1/2) \, {\bf P} \cdot {\bf \Sigma} \notag  
\\
&& \qquad\qquad
+(\nabla {\bf u}) \cdot {\bf P} - (8/\eps^2)D_\eps \lambda_\Sigma\boldsymbol
\Sigma \cdot {\bf P} 
\label{e:P-equation} \\
&& \notag\qquad\qquad
+ \ga{\bf P} - \gb |{\bf P}|^2{\bf P}  + \Gamma_0 \nabla^2{\bf P}  
-\Gamma_2 \nabla^4 {\bf P },
\ee
where $\gl_\mathrm{P*} =(\mu/\mu_*)\gl_\mathrm{P}$ and the effective
orientation-pressure field is given by
\be
\hat{p}_*=
 \hat{p}+v_0
 \left(
-\f{\gl_\mathrm{P}}{2} |{\bf P}|^2  + 
\f{\lambda_\Sigma}{2{\mu}_*}   {p}_*\right).
\ee
The remaining parameters in Eq.~\eqref{e:P-equation} are obtained as 
\be
\Gamma_0&=&
(D + D_\eps) -\f{\lambda_\Sigma v_0 f_0 \phi}{2{\mu}_*
A\ell}\gamma_2,
\qquad \ga=\f{4D_\eps}{\eps^2}-\f{1}{\tau},
\notag
\\
\Gamma_2&=&
\f{\lambda_\Sigma v_0 f_0\phi}{2{\mu}_* A \ell} \gamma_4-  
\f{\eps^2 D_\eps}{24}
\csp\quad
\gb= \f{4D_\eps}{\eps^2} \lambda_\mathrm{P}.
\label{e:P-parameters}
\ee
Equation~\eqref{e:P-equation} is structurally similar to the Toner-Tu equation \cite{Toner_Tu}
with the significant difference that the
 \lq diffusion\rq~parameter $\Gamma_0$ can become negative when the
volume fraction $\phi$ and active power $v_0f_0$ become sufficiently large, as
proposed earlier on purely phenomenological grounds~\cite{NJP,PRL,PNAS}.  For
$\Gamma_0<0$, Eq.~\eqref{e:P-equation} predicts a transition to mesoscale 
turbulence, as observed in dense {\em B.~subtilis} suspensions \cite{Aranson_collective_motion,PNAS,PRL}.

\begin{figure}[t]
\begin{center}
\includegraphics[width=\columnwidth]{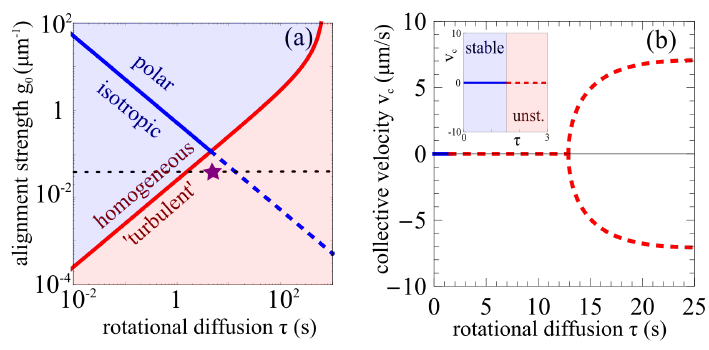} 
\caption{(color online) Bifurcation analysis. (a)~State diagram for  rod-like pusher obtained by a
linear stability analysis of Eq.~\eqref{FEQ} for typical \textit{B. subtilis} parameters (see text) and $v_0 = 10\, \mu$m/s. The red 
line demarcates the transition to mesoscale turbulence. The blue
line signals the transition between disorder and polar order. The purple star
indicates the parameters used in simulations. (b) Bifurcation diagram of the collective velocity $v_c = v_0 \sqrt{\alpha/\beta}$ for rod-like
pushers along the dotted
black line ($g_0=0.04\, \mu$m$^{-1}$) in (a). Red dashed lines depict unstable branches, whereas  blue solid lines depict stable branches. Inset: Zoom to $\tau \in [0,3]$s.}
\label{FigBif}
\end{center}
\end{figure}

\paragraph{Parameters.--}
 The coefficients in Eqs.~\eqref{e:stokes_new}-\eqref{e:P-parameters}
can be directly estimated from experiments~\cite{Wolgemuth,PRL,PNAS,Aranson_collective_motion,Sokolov09}:
 In our simulations, we consider parameters for
{\em B. subtilis} bacteria (cell length $\sim 5\, \mu$m and diameter $d=0.7\,
\mu$m) at  high volume fractions $\phi
\sim 0.4$ \cite{PNAS,Wolgemuth}, assuming an effective dipole length $\ell =b_r=
5\, \mu$m~(Fig.~\ref{sketch}) and for the
projected 2D area $A\approx d \ell$.
The typical force $f_0$ exerted by a single microswimmer on the 
surrounding fluid can be estimated as $ f_0  \approx  2 \pi \mu_0 \ell  v_0 $
\cite{Wolgemuth},  with typical  bacterial  self-propulsion speed  $v_0\in[
1,50] \,\mu$m/s \cite{Aranson_collective_motion}. In the collision-dominated
high-density regime relevant to our study, translational Brownian motion is
negligible, $D\ll D_\eps$, and we set $D=0$ in our simulations. We further 
assume that steric short-range interactions occur predominantly on the length
scale of cell-body, $\epsilon = 3\,\mu$m.   After fixing the above parameter values, 
we can analyze how changes of the rotational
relaxation time $\tau$, alignment strength $g_0$, and swimming activity
$v_0$ affect the collective dynamics, by exploring the range  
$\tau\sim 0.01...10^4\,$s, $g_0=0.001...10^3 \,  \mu$m$^{-1}$, and $v_0=1\ldots 40\mu$m/s~(Fig.~\ref{FigBif}).

\begin{figure}[b]
\begin{center}
\includegraphics[width=\columnwidth]{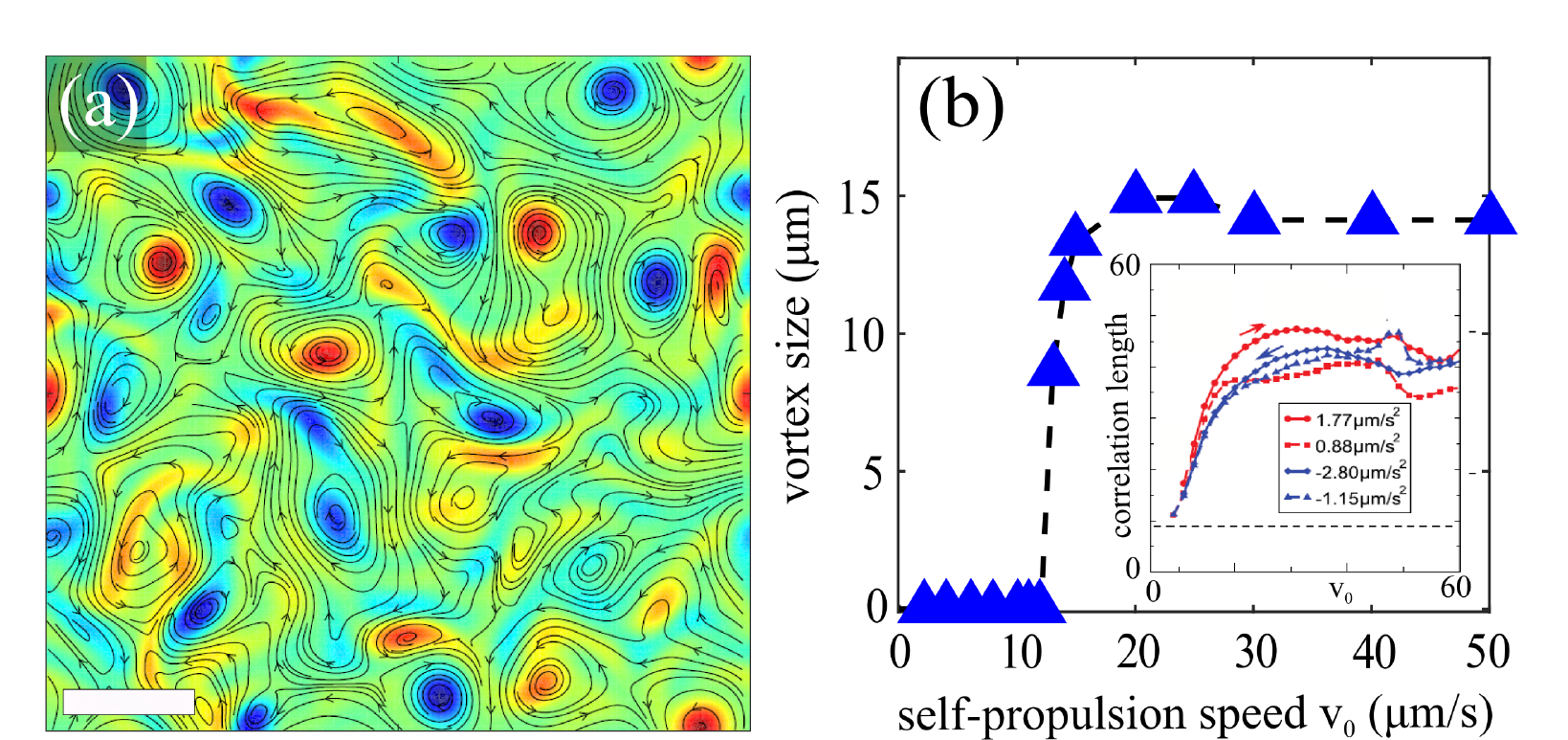} 
\caption{(color online) (a)  Representative snapshot of the effective velocity field $\bf{w}$
 from a simulation with typical \textit{B. subtilis} parameters (see text), $v_0= 20 \mu$ms$^{-1}$ and $g_0=0.04\,\mu$m$^{-1}$. Scale bar corresponds to $20\mu$m and color coding indicates vorticity normalized by the maximum. 
(b) In dense suspensions, the characteristic vortex size approaches a constant
value at high large activity.
This prediction agrees qualitatively with recent
measurements~\cite{Aranson_collective_motion} shown in the inset.
\label{FigSim}
}
\end{center}
\end{figure}
\paragraph{Bifurcation diagram.--}
The field equations~\eqref{e:stokes_new} and \eqref{FEQ} have two fixed points: the disordered
state \mbox{(${\bf u} ={\bf 0}, \bf{P} = \bf{0}$)} and the polar ordered state \mbox{(${\bf u} ={\bf 0}, \bf{P} \ne \bf{0}$)}. Upon varying $\tau$ and $g_0$, these homogeneous states become unstable when the
alignment strength becomes subcritical relative to the rotational noise [red
line in Fig.~\ref{FigBif}(a)]. Conversely, strong alignment stabilizes the homogeneous polar state. Defining $\tau$ as control parameter
and the collective velocity $v_c = v_0 \sqrt{\alpha/\beta}$ as an order
parameter, the dotted black line of the state diagram yields the
bifurcation diagram of Fig.~\ref{FigBif}(b).  Upon linearizing Eq.~\eqref{e:P-equation} about the isotropic state,
the typical vortex length follows from the most unstable mode, which has
wavelength $\Lambda \sim 2 \pi \sqrt{ 2 \Gamma_2/ (-\Gamma_0)}$.  As evident
from the explicit expressions for $\Gamma_0$ and $\Gamma_2$ in
Eq.~\eqref{e:P-parameters}, this vortex scale is set by the competition between 
hydrodynamic flows, 
steric alignment interactions and  activity. In particular, in the limit of
strong self-propulsion and high concentrations, the theoretically predicted
vortex size approaches a constant value in agreement with recent
experiments~\cite{Aranson_collective_motion,PNAS}.

\paragraph{Simulations vs. experiment.--}
To study the full nonlinear behavior, we solved Eqs.~\eqref{e:stokes_new}
and~\eqref{e:P-equation}  numerically with a pseudo-spectral code that combines
anti-aliasing with an operator splitting technique \cite{PNAS}.  Simulations
were performed using $128\times 128$ grid points for an 
area of  $101 \times 101\, \mu$m$^2$ and time steps of $dt = 10^{-3}$s,
respectively, for a total simulation time in the range
$[500,1000]$s. For typical \textit{B. subtilis} parameters and
$\tau=4.5$s~, $g_0=0.04 \mu$m$^{-1}$, we obtain flow structures that
agree with recently measured flow fields [Fig.~\ref{FigSim}(a)]. Moreover,  the
numerically measured vortex scale, obtained from the minimum of the velocity
correlation function, approaches a constant value  at large activity, in
qualitative agreement with recent measurements~\cite{Aranson_collective_motion}, see  Fig.~\ref{FigSim}(b).

\paragraph{Conclusions.--}
We presented a systematic derivation of higher-than-second-order
hydrodynamic equations from a generic microswimmer model. The resulting field
theory explains simultaneously a number of recent 
experimental observations, including the reduction (enhancement) of viscosity in
pusher (puller) suspensions and  the emergence of a finite characteristic vortex
size in dense active fluids from the interplay of fluid-dynamical and steric
interactions. Generally,
this work shows that higher-order theories of active suspensions~\cite{PNAS,PRL} arise naturally
if one adopts self-consistent closure conditions. Future work may focus on
generalizing the above approach to active nematics~\cite{2012Sanchez_Nature}.

\begin{acknowledgments}
We are grateful to  L. Schimansky-Geier,  R. Gro{\ss}mann, P. Romanczuk  and C. Marchetti for
discussions.  This work was supported by the Deutsche
Forschungsgemeinschaft through GRK 1558 and SFB 910 (S.H., S.H.L.K., M.B.),
 by an MIT Solomon Buchsbaum Fund Award ~(J.D.) and an Alfred P. Sloan Research
Fellowship~(J.D.). 
\end{acknowledgments}

\bibliographystyle{apsrev4-1}
\bibliography{PRL_submitted}

\end{document}